\documentclass[a4paper,11pt]{article}

\pdfoutput=1

\usepackage{amsmath}
\usepackage{amsfonts}
\usepackage{amssymb}
\usepackage{comment}
\usepackage[colorlinks,citecolor=blue,urlcolor=blue,bookmarks=false,hypertexnames=true]{hyperref}
\usepackage{cite}

\usepackage{graphicx, physics, subcaption}
\usepackage{verbatim }

\numberwithin{equation}{section}

\usepackage{color}

\usepackage{amsmath, amssymb, graphics, epsfig, graphicx}
\usepackage{epsf}
\usepackage{epstopdf}
\usepackage {amssymb}
\newcommand{\nc}{\newcommand}

\newcommand{\calR}{{\cal{R}}}

\def\bfk{{\bf k}}
\def\bfq{{\bf q}}

\nc{\cN}{ {\cal{N}} }

\nc{\ba}{\begin{eqnarray}}
\nc{\ea}{\end{eqnarray}}

\begin{document}

\vspace{5mm}
\vspace{0.5cm}
\begin{center}

\def\thefootnote{\fnsymbol{footnote}}

{\bf\large {{Trispectrum in Extended USR Model  with Transition to SR } }}
\\[0.5cm]

{ Hassan Firouzjahi$^a$ $\footnote{firouz@ipm.ir}$, Amin Nassiri-Rad$^b$ $\footnote{amin.nassiriraad@kntu.ac.ir}$ }   \\[0.5cm]
{\small \textit{$^a$ School of Astronomy, Institute for Research in Fundamental Sciences (IPM) \\ P.~O.~Box 19395-5746, Tehran, Iran}\\
\small \textit{$^b$ Department of physics, K.N Toosi University of Technology, P.O. Box 15875-4416, Tehran, Iran}
}

\end{center}

\vspace{.8cm}

\hrule 
\vspace{0.3cm}


\begin{abstract}
We study the trispectrum in a two-phase USR-SR setup of inflation in which the USR stage is extended in the initial phase of inflation while the second stage of inflation proceeds via a slow-roll phase.  A key role is played by the sharpness parameter which controls how quickly the system reaches the final attractor phase after the USR stage. We employ both $\delta N$ and in-in formalisms and calculate trispectrum and the corresponding dimensionless parameters $g_{NL}$ and $\tau_{NL}$. 
We show that both approaches yield the same results and study the  
shapes of trispectrum in various configurations. 
It is shown that  the maximum value of trispectrum occurs in the setup with an infinitely sharp transition to the attractor phase while much of trispectrum is washed out in the opposite limit of a mild transition.

\end{abstract}
\vspace{0.5cm} \hrule
\def\thefootnote{\arabic{footnote}}
\setcounter{footnote}{0}
\newpage

\section{Introduction}
Models of ultra-slow-roll (USR) inflation \cite{Kinney:2005vj} have been studied extensively in the literature. The USR setup in its simplest realization is a short period  of  inflation during which the inflaton potential is exactly flat with $V=V_0$ and  while inflaton is rolling,  its kinetic energy is exponentially diluted.  This short period is terminated either abruptly or smoothly  where the constant potential is glued to a new segment  of potential which can support an extended period of slow-roll (SR) inflation. 
There are two main reasons for the interests in USR models. Originally \cite{Namjoo:2012aa, Chen:2013aj}, it was studied as a counter example  of the single field models which can violate the Maldacena non-Gaussianity consistency condition \cite{Maldacena:2002vr}. The prime reason for the violation of the consistency condition is the non-conservation of the curvature perturbation on superhorizon scales which itself is a direct consequence of the exponential dilution of the inflaton velocity during the USR phase.

The second reason for the interests in USR setup was that it can be engineered to generate primordial black holes (PBHs) which can be a good candidate for all or part of dark matter \cite{ Garcia-Bellido:2017mdw, Germani:2017bcs, Biagetti:2018pjj}, (see also \cite{Ivanov:1994pa} for earlier work) and  
\cite{Khlopov:2008qy, Ozsoy:2023ryl, Byrnes:2021jka, Escriva:2022duf, Pi:2024jwt} for more reviews. In these setups, a short period of USR is sandwiched between two long periods of SR inflation. The first period is where the long CMB modes leave the horizon while during the intermediate USR phase the curvature perturbation grows exponentially to generate the PBHs of right mass scales and amplitudes. Finally, inflation ends when the USR phase is followed by the second SR phase and reheating.  This setup is known as the three phase SR-USR-SR model.

In this work, similar to original studies such as \cite{Namjoo:2012aa, Chen:2013aj}, we consider a two-phase setup in which the first stage is a USR  phase during which the long CMB modes are generated. This phase may be extended in the past expansion history so for this reason we call it an extended phase, in the sense that it is not sandwiched between SR  phases. Since the curvature perturbation grows exponentially during the USR phase, the USR setup becomes non-perturabtive so we have to terminate it by a SR attractor phase. 
The bispectrum (the three-point function) in this 
setup was studied extensively  see for example \cite{Namjoo:2012aa, Martin:2012pe, Chen:2013aj, Morse:2018kda, Lin:2019fcz, Dimopoulos:2017ged, Chen:2013eea, Akhshik:2015nfa, Akhshik:2015rwa, Mooij:2015yka, Bravo:2017wyw, Finelli:2017fml, Passaglia:2018ixg, Pi:2022ysn, Ozsoy:2021pws,  Firouzjahi:2023xke, Namjoo:2023rhq, Namjoo:2024ufv, Cai:2018dkf}. In particular, in the simple setup with the standard kinetic energy where the USR phase is followed immediately by an attractor SR phase, it is shown \cite{Namjoo:2012aa} that  the amplitude of non-Gaussianity is  $f_{NL}=\frac{5}{2}$. Having said this, we are not aware of works which have studied  the trispectrum (four-point functions) in this USR-SR setup systematically. The goal of this work is to study trispectrum in this setup  in some details.  For this purpose, we employ both  $\delta N$  and  in-in formalisms. Each method has its own advantages and we confirm that both methods yield the same results for the shapes of the trispectrum.

\section{The Model}
\label{model}

In this section we present our setup. It is a two-stage model of single field  inflation involving the inflaton field $\phi$ with the potential $V(\phi)$. 
The first stage is an extended USR phase of inflation for the period $\tau < \tau_e$ in which $\tau$ is the conformal time and  $\tau_e$ is the end of the USR phase. The second stage during $\tau_e< \tau <\tau_0$ is a SR phase of inflation with $\tau_0\rightarrow 0$ representing the time of end of inflation.  
Without loss of generality, we assume that $\phi$ is 
monotonically  decreasing so  the first stage corresponds to $\phi > \phi_e$ while during the second stage $\phi_e < \phi < \phi_f$, 
in which $\phi_e$ is the value of the field at the end of USR while $\phi_f$ represents the value of field at the end of inflation. 
The transition from the USR phase to SR phase takes place instantaneously at $\tau_e$. This is mainly for analytical purposes and in a real situation one expects that the transition from the USR phase to the SR phase to be smooth. However, demanding the potential to be smooth will make the theoretical analysis intractable and a full numerical analysis will be required. During the USR phase the inflaton potential is constant, $V(\phi)= V_0$, while during the second stage it supports a SR dynamics with the first and second derivatives of the potential to be non-zero. As in conventional SR setups,  inflation ends when the SR conditions are violated in the final stage followed by reheating. We assume that the first stage is long enough and the CMB modes leave the horizon during the USR stage.   For additional reviews on different aspects of USR setup see \cite{Kinney:2005vj, Namjoo:2012aa, Martin:2012pe, Chen:2013aj, Morse:2018kda, Lin:2019fcz, Dimopoulos:2017ged, Chen:2013eea, Akhshik:2015nfa, Akhshik:2015rwa, Mooij:2015yka, Bravo:2017wyw, Finelli:2017fml, Passaglia:2018ixg, Pi:2022ysn, Ozsoy:2021pws,  Firouzjahi:2023xke, Namjoo:2023rhq, Namjoo:2024ufv, Cai:2018dkf}.

Considering the FLRW metric,
\ba
ds^2 = -dt^2 + a(t)^2 d{\bf x}^2 \, ,
\ea 
the dynamics of the background during the USR phase is given by, 
\ba
\label{USR-back}
\ddot \phi(t) + 3 H \dot \phi(t)=0\, , \quad \quad 3 M_P^2 H^2 \simeq V_0, 
\ea
where $M_P$ is the reduced Planck mass and $H=\dot a(t)/a(t)$ is the Hubble expansion rate during inflation. A key feature of the USR phase is that 
since the potential is flat, $\dot \phi$ falls off exponentially, $\dot \phi \propto a(t)^{-3}$ and correspondingly the first slow-roll parameter $\epsilon\equiv -\dot H/H^2$ falls off like $\epsilon \propto a(t)^{-6}$. As a result, the second slow-roll parameter 
$\eta\equiv \dot \epsilon/H \epsilon$ is large with $\eta \simeq -6$ \cite{Kinney:2005vj}.   With this description, the evolution of $\epsilon(\tau)$ during the USR phase can be written as, 
\ba
 \epsilon(\tau) = \epsilon_e \big( \frac{\tau}{\tau_e} \big)^6 \, , \quad \quad (\tau < \tau_e) \, ,
 \ea
in which $\epsilon_e= \epsilon(\tau_e)$.

During the follow up SR stage, the potential can be approximated by its first and second slow-roll parameters as follows,
\ba
V(\phi) \simeq V(\phi_e) + \frac{\sqrt{2 \epsilon_V}}{M_P} V(\phi_e)  (\phi -\phi_e) + \frac{\eta_V}{2 M_P^2} V(\phi_e) (\phi -\phi_e)^2  \, ,
\ea
in which the SR parameters $\epsilon_V$ and $\eta_V$ are defined with respect to the potential at $\phi_e^+$ as follows, 
\ba
\epsilon_V \equiv \frac{M_P^2}{2}\frac{V_{,\phi}^2}{V^2}, \quad \quad
\eta_V\equiv M_P^2 \frac{V_{,\phi \phi}}{V} , \quad \quad (\phi= \phi_e^+) \, .
\ea
The potential is continuous at $\phi_e$ but it has a kink at $\phi_e$ since we assume  $\epsilon_V \neq 0$. However, to simplify the analysis further, we assume that $\eta_V =0$. This is the sharp transition limit $\sqrt{2 \epsilon_V} \gg \eta_V $ as discussed in \cite{Cai:2018dkf}. However, this does not bring restrictions in our analysis and one can consider a more general limit where  $\eta_V \neq 0$ but this brings more complications in theoretical analysis. 

Using the number of e-folds $dN= H dt$ as the clock, and considering the above discussions in mind,  the dynamics of the background in the SR phase is given by ,
 \ba
 \label{KG-SR}
 \frac{d^2 \phi}{ d N^2} + 3 \frac{d \phi}{d N} + 3 M_P \sqrt{2 \epsilon_V} \simeq 0 \, ,
 \quad \quad 3 M_P^2 H^2 \simeq V(\phi_e) \, .
 \ea
Let us assume  the time of transition from the USR to SR to be at $N=0$. Requiring that the field and its first derivative to be continuous  at $N=0$, the solution is given by, 
\ba
M_P^{-1}\phi(N)= \frac{C_1}{3} e^{-3 N} + \frac{h}{6}  \sqrt{2 \epsilon_e} N + C_2 \, ,
\ea
with,
 \ba
 C_1=  \sqrt{2 \epsilon_e} (1 + \frac{h}{6} ) \, , \quad \quad 
 C_2 = M_P^{-1} \phi_e - \frac{ \sqrt{2 \epsilon_e}}{3} (1 + \frac{h}{6} ) \, .
 \ea
Here we have defined the sharpness parameter $h$ via \cite{Cai:2018dkf},  
\ba
\label{h-def}
h\equiv \frac{6 \sqrt{2 \epsilon_V} }{\dot \phi(t_e)} M_P H = -6 \sqrt{\frac{\epsilon_V}{\epsilon_e}} \, .
\ea

A sharp transition corresponds to the case $|h| >1$ in which the mode function quickly approaches its final attractor value. In the case of extreme sharp transition corresponding to $| h| \rightarrow \infty$, the mode function freezes immediately after the transition which is the limit considered originally in \cite{Namjoo:2012aa, Chen:2013aj}. A particular case of sharp transition is where $h=-6$ in which $\epsilon_V = \epsilon_e$ as considered for example in \cite{Kristiano:2022maq, Kristiano:2023scm}. This is called the instant transition, but even in this case the mode function keeps evolving for some time after the USR transition until it 
assumes its final attractor value. 
On the other hand, a mild transition corresponds to the situation where $|h| \ll 1$ during which the mode function keeps evolving towards the end of inflation. In this situation, one should keep track of the evolution of the mode function. As shown in \cite{Cai:2018dkf}, during the mild transition much of the non-Gaussianity accumulated in the USR phase is washed out while in an extreme sharp transition   the amplitude of non-Gaussianity remains mostly intact with $f_{NL} =\frac{5}{2}$ \cite{Namjoo:2012aa, Chen:2013aj}. However, in a general case, $f_{NL}$ depends on $h$ which we provide the corresponding formula later on. In the limit of a mild transition where 
$h$ is as small as the SR parameters, one should keep track of the SR effects as well. In this work, to simplify the analysis and in order to perform the calculations analytically, we consider the limit of sharp transition with $|h| >1$.

The evolution of the first and the second SR parameters in the SR stage ($N>0$) are given by, 
\ba
\label{ep-N}
\epsilon(\tau)= \epsilon_e  \Big(\frac{h}{6} - (1+ \frac{h}{6} ) \big(\frac{\tau}{\tau_e} \big)^3 \Big)^{2} \, ,
\ea
and
\ba
\label{eta-N}
\eta(\tau) = -\frac{6 (6+h)}{(6+h) - h   \big(\frac{\tau_e}{\tau} \big)^3} \, .
\ea 
Note that $\epsilon(\tau)$ and $\eta(\tau)$ in the above expressions are defined with the evolution  of $H$ and are different from $\epsilon_V$ and $\eta_V$ which are defined with respect to derivatives of potential and are  nearly constants  in the SR limit. However, $\epsilon(\tau_0) \simeq \epsilon_V$ as the latter is the SR parameter in the attractor limit. 

From the structure of $\eta(\tau)$ in Eq. (\ref{eta-N}) we see that near 
$\tau=\tau_e^+$ it is approximately given by $\eta \simeq -6 -h$ while during the USR phase, as mentioned before,  it is $\eta=-6$. Therefore, we can approximate 
the evolution of $\eta$ near the point of transition via \cite{Cai:2018dkf}, 
\ba
\eta = -6 - h \theta(\tau -\tau_e) \quad \quad  \tau_e^- < \tau < \tau_e^+ \, ,
\ea
in which $\theta(x)$ is the step function. This yields to the following formula for the derivative  of $\eta$, 
\ba
\label{eta-jump}
\frac{d \eta}{d \tau} = - h \delta (\tau -\tau_e)  \, ,  \quad \quad  \tau_e^- < \tau < \tau_e^+ \, .
\ea
This indicates that $\eta$ has a jump at the point of transition which is controlled by the parameter $h$. As we shall see, $\eta'$ induces a local source in the interaction Hamiltonian which plays crucial roles in our bispectrum and trispectrum analysis.

Having described the background, we look at the mode function of the comoving
curvature perturbation $\calR$. Going to the Fourier space, $\calR$ is expanded as follows, 
\ba
\calR({\bf x}, t) = \int \frac{d^3 k}{(2\pi)^3} e^{i {\bf k}\cdot {\bf x}} \hat\calR_{\bf k}(t) \, ,
\ea
 in which  $\hat\calR_{\bf k}(t)= \calR_k(t) a_{\bf k} + \calR^*_k(t) a_{-\bf k}^\dagger$. Here $a_{\bf k}$ and $a_{\bf k}^\dagger$ are the annihilation and creation operators respectively which satisfy the usual  commutation relations $[ a_{\bf k}, a^\dagger_{-\bf k'} ] = ( 2 \pi)^3 \delta (  {\bf k} + {\bf k'}) $.

Starting with the Bunch-Davies (Minkowski) initial condition for the modes deep inside the horizon, the mode function during the USR stage is given by,
\begin{equation}
\label{onephase}
\calR_{k} =  \frac{H}{ M_P\sqrt{4 \epsilon_e k^3}}  \bigg( \frac{\tau_e}{\tau} \bigg)^3
( 1+ i k \tau) e^{- i k \tau} \,  \quad \quad (\tau< \tau_e) .
\end{equation}
During  the final SR phase, after imposing the continuity of $\calR$ and $\calR'$ at $\tau=\tau_e$,  we obtain \cite{Cai:2018dkf, Firouzjahi:2023aum},
\begin{equation}
\label{R-alpha-beta}
\calR_{k} =  \frac{H}{ M_P\sqrt{4 \epsilon(\tau) k^3}} \Big[ \alpha_k ( 1+ i k \tau) e^{- i k \tau}  + \beta_k ( 1- i k \tau) e^{ i k \tau}  \Big]  \,   \quad \quad (\tau_e< \tau< \tau_0)\, ,
\end{equation}
where $\epsilon(\tau)$ is given in Eq. (\ref{ep-N}) and 
the constants $\alpha_k$ and $\beta_k$ are given by, 
\begin{equation}
\label{alpha-beta}
\alpha_k = 1 + \frac{ i h}{ 4 k^3 \tau_e^3} ( 1 + k^2 \tau_e^2) \,,  \quad \quad
\beta_k= -\frac{i h}{ 4 k^3 \tau_e^3 } {( 1+ i k \tau_e)^2} e^{- 2 i k \tau_e} \, .
\end{equation}
In the next section we compute the trispectrum using the mode function given above. 

Having obtained the mode function, let us look at the power spectrum  at the end of inflation $\tau=\tau_0 \rightarrow 0$. Using the expression for $\calR(\tau)$ given in Eq. (\ref{R-alpha-beta}), and noting that $\epsilon(\tau_0)\simeq \epsilon_V$ in the SR limit,  we obtain,
\ba
\label{PR-USR}
P_\calR(k, \tau_0) = \big| \calR_k(\tau_0)  \big|^2 = \frac{(h-6)^2}{ h^2}  \Big(\frac{H^2}{4 k^3 M_P^2 \epsilon_e} \Big) = \Big(  1- \frac{h}{6}
\Big)^2  \Big(\frac{H^2}{4 k^3M_P^2 \epsilon_V} \Big) \, .
\ea 
In the limit of extreme sharp transition with $h \rightarrow -\infty$, we obtain the expected result that $P_\calR(\tau_0) = \Big(\frac{H^2}{ 4 k^3M_P^2 \epsilon_e} \Big)= P_\calR(\tau_e)$ as the mode function freezes immediately after the USR phase. In the case of an instant transition with  $h=-6$, we obtain the curios result that  $P_\calR(\tau_0)=  4 P_\calR(\tau_e)$ so the power at the end of inflation is larger by a factor 4 compared to its value at the end of USR. This is a direct realization of the fact that the mode function keeps evolving after the USR phase.

A key feature of the USR setup is that the curvature perturbation is not frozen on superhorizon scales. More specifically, since $\epsilon \propto a^{-6}$, from Eq. (\ref{onephase}) we see that $\calR(\tau)$ grows like $a(\tau)^3$. As a result, in the USR model the would-be decaying mode of $\calR$ is actually the growing mode, causing the violation of the celebrated Maldacena non-Gaussianity consistency condition in single field scenarios \cite{Maldacena:2002vr}.
The bispectrum in the current setup with the sharpness parameter $h$ is 
calculated in \cite{Cai:2018dkf, Firouzjahi:2023aum} with the amplitude of non-Gaussianity $f_{NL}$ given by (in our limit where $\eta_V\rightarrow 0$),
\ba
\label{fNL}
f_{NL} = \frac{5 h^2}{ 2 (h-6)^2} \, .
\ea
In the limit of extreme sharp transition $h\rightarrow -\infty$ we recover the original result \cite{Namjoo:2012aa, Chen:2013aj} $f_{NL} =\frac{5}{2}$. However, as noticed in  \cite{Cai:2018dkf}, if the transition is mild with $|h| \ll1$, much of non-Gaussianity is washed out during the USR phase with $f_{NL}\sim  h \eta_V$. However, even in this case, Maldacena's non-Gaussianity consistency condition is still violated.

After reviewing our setup, we are ready to calculate the trispectrum in this setup. For earlier works on trispectrum mainly in the context of multi field inflation or $P(X, \phi)$ models of inflation see for example 
\cite{Seery:2006vu, Jarnhus:2007ia, Arroja:2008ga, Arroja:2009pd, Mizuno:2009cv, Mizuno:2009mv, Izumi:2011di, Chen:2009bc, Chen:2018sce, Chen:2009zp, Renaux-Petel:2009jdf, Gao:2009at, Gao:2009gd, Gao:2010xk, Izumi:2010wm, Bartolo:2009kg, Bartolo:2010di, Leblond:2010yq, Sheikhahmadi:2019xkx}. 
We calculate the trispectrum from both $\delta N$ and in-in formalisms and examine various consistencies between these two approaches. In our analysis below, we assume that all four modes leave the horizon during the USR stage so they are all superhorizon at the time $\tau_e$. Of course, one can consider a general case where some modes leave the horizon during the SR stage. In this case, more shapes of trispectrum beyond what we study here will be generated.  

\section{Trispectrum from $\delta N$ Formalism}
\label{DeltaN}

In this section we calculate the trispectrum using $\delta N$ formalism which is proved to be easier than the in-in approach which we postpone to the next section. 

The $\delta N$ formalism is a powerful tool to study cosmological perturbations non-linearly \cite{Sasaki:1995aw, Sasaki:1998ug,  Wands:2000dp, Lyth:2004gb, Lyth:2005fi, Abolhasani:2019cqw}. It relies on the separate universe picture in which it is assumed that the nearby Hubble size patches evolve independently as separate FLRW backgrounds with different initial conditions which are inherited from initial horizon size perturbations. The comoving curvature perturbation is related to the difference in the  number of e-folds  between two nearby 
patches. The number of e-folds $N$ is counted between an initial and a final hypersurfaces in which  the initial hypersurface is spatially flat while the final hypersurface is the surface of constant energy density.  
To employ the $\delta N$ formalism, one has to solve $N$ as a function of the background field and its velocity, $N= N(\phi, \dot \phi)$. Having obtained $N(\phi, \dot \phi)$ one can expend it perturbatively to any order to calculate the power spectrum, bispectrum, trispectrum etc. More schematically (and neglecting $\delta \dot \phi$ which vanishes on superhorizon scales),   
\ba
\label{delta N}
\delta N= N' \delta \phi + \frac{N''}{2} \delta \phi^2 + \frac{N'''}{3!} \delta \phi^3+...
\ea
in which a prime on $N$ means the derivative with respect to $\phi$, i.e. 
$N'= N_{, \phi}$ and so on. Eq. (\ref{delta N}) is based on the perturbative expansion of $\delta N$ assuming the system is perturbative. However, a great advantage of $\delta N$ formalism is that it allows for non-perturbative analysis as well, in situations like the tail of the  PBHs distributions which are rare events and non-perturbative, requiring non-linear analysis \cite{Hooshangi:2021ubn, Cai:2021zsp, Cai:2022erk, Kawaguchi:2023mgk}

The comoving curvature perturbation $\calR$ is related to $\delta N$ via 
$\calR= -\delta N$. Using the leading term in the expansion (\ref{delta N}), this yields to the power spectrum,
\ba
P_\calR(k) = (N')^2 P_{\delta \phi} = (N')^2 \frac{H^2}{4 k^3}  \, . 
\ea
On the other hand, the bispectrum is related to the three-point function via,
\ba
\langle \calR_{\bfk_1} \calR_{\bfk_2} \calR_{\bfk_3}\rangle = (2 \pi)^3 \delta^3(\bfk_1+ \bfk_2+ \bfk_3)  B_\calR(k_1, k_2, k_3)  \, .
\ea
To calculate the bispectrum, we need to consider the second order perturbations in  Eq. (\ref{delta N}). As the perturbations $\delta \phi$ are Gaussian and free on superhorizon scales, only the local-type non-Gaussianity is generated which is also consistent with the expansion Eq. (\ref{delta N}) where the curvature perturbation at the quadratic order is the square of the first order perturbation. This in turn yields to the local-shape bispectrum  \cite{Byrnes:2006vq}, 
\ba
 B_\calR(k_1, k_2, k_3)  = \frac{6}{5} f_{NL} 
 \Big[P_\calR(k_1) P_\calR(k_2)  + P_\calR(k_1) P_\calR(k_3)+ P_\calR(k_2) P_\calR(k_3) \Big] \, ,
\ea
yielding to,
\ba
\label{fNL-eq}
f_{NL}= \frac{5}{6} \frac{N''}{(N')^2} \, .
\ea

The trispectrum is related to the four-point function as follows,
\ba
\label{trispectrum}
\langle \calR_{\bfk_1} \calR_{\bfk_2} \calR_{\bfk_3} \calR_{\bfk_4} \rangle = (2 \pi)^3 \delta^3(\bfk_1+ \bfk_2+ \bfk_3+ \bfk_4)  T_\calR(k_1, k_2, k_3, k_4) \, .
\ea
Due to the non-linear structure of the convolutions integrals involved in Fourier space, we have two distinct local shapes in trispectrum, parameterized as follows  \cite{Byrnes:2006vq}, 
\ba
\label{TR}
T_\calR(k_1, k_2, k_3, k_4) &=& \tau_{NL} \big[ P_\calR(k_{13}) P_\calR(k_3) P_\calR(k_4) + (11~ \text{perms}) \big] \nonumber\\
&+& \frac{54}{25} g_{NL} \big[ P_\calR(k_{2}) P_\calR(k_3) P_\calR(k_4)
+ (3~ \text{perms}) \big] \, ,
\ea
in which  $k_{ij}\equiv | \bfk_i+\bfk_j |$. 

Here, $\tau_{NL}$ and $g_{NL}$ are two new parameters which describe local-shape trispectrum given by,
\ba
\label{gNL-tauNL}
g_{NL} = \frac{25}{54} \frac{N'''}{(N')^3} , \quad \quad
\tau_{NL} = \frac{(N'')^2}{(N')^4} \, .
\ea
An immediate conclusion from the expressions for $\tau_{NL}$ and $f_{NL}$
in Eqs. (\ref{gNL-tauNL}) and (\ref{fNL-eq}) is that $\tau_{NL}= \frac{36}{25} f_{NL}^2$ which is the special case of the Suyama-Yamaguchi inequality \cite{Suyama:2007bg} $\tau_{NL} \ge \frac{36}{25} f_{NL}^2$ where the equality is valid for any single field model of inflation in the tree level. 

The above discussions are general, valid to any single field scenario as long as $\delta \phi$ is treated as a Gaussian free field. This also means that we work in tree level in which the loop corrections in bispectrum and trispectrum are neglected since in the presence of loops, the $\delta \phi$ perturbations may not be considered as a free and Gaussian field.\footnote{For the effects of loop corrections in initial $\delta \phi$ amplitude employed in $\delta N$ formalism see \cite{Nassiri-Rad:2025dsa}.} 

To calculate the amplitudes of non-Gaussianities such as $f_{NL}, g_{NL}$ and  $\tau_{NL}$, all is left is to solve for $N(\phi, \dot \phi)$ and calculate its derivatives. Note that since the system is in non-attractor phase during the USR stage, $\phi$ and $\dot \phi$ are independent variables so $N(\phi, \dot \phi)$ should be calculated in the phase space. This is the key difference in employing $\delta N$ formalism in USR setup 
compared to conventional  SR setups. 
The analysis to calculate $N(\phi, \dot \phi)$ in our two-stage USR-SR model of inflation were presented in \cite{Cai:2018dkf}, see also \cite{Firouzjahi:2023ahg} for the corresponding $\delta N$ analysis of the three-stage setup SR-USR-SR. Here we briefly review the main steps to calculate $N(\phi, \dot \phi)$.

During the USR phase, the scalar field equation (\ref{USR-back}) is 
cast into,
\ba
\pi'(N) + 3 \pi(N) =0\, , \quad \quad \pi\equiv \frac{d \phi}{ d N} \, ,
\ea
yielding to the solution,
\ba
\label{pi-N}
\pi(N)= \pi_e e^{-3 N} \, ,
\ea
where we have set the time of transition from the USR phase to SR  to be at  $N=0$. Therefore, during the USR period $N<0$ while during the final SR stage $N>0$. 
Also, note that $\pi_e\equiv \pi(N=0)$ represents the velocity of the field at the point of the transition. 

Eq. (\ref{pi-N}) can be solved easily for $\phi(N)$, yielding,
\ba
\label{phi-USR}
\phi(N)= \phi_e+ \frac{\pi_e}{3} \big( 1- e^{-3 N} \big) \, .
\ea
The combination of Eqs. (\ref{phi-USR}) and (\ref{pi-N}) yields the following constraint, 
\ba
\label{constraint}
\pi_e= \pi + 3 (\phi - \phi_e) \, .
\ea
Finally, solving for $N$ from Eq. (\ref{phi-USR}) and using the constraint (\ref{constraint}), yields,
\ba
\label{N-USR}
N_{\text{USR}}= \frac{1}{3} \ln \Big[1+ \frac{3 (\phi - \phi_e)}{\pi}
\Big]\, .
\ea
Here, $N_{\text{USR}}$ means the number of e-folds during the USR 
phase,  which takes the field to move from the initial positions $(\phi, \pi)$ in the phase space to the end point in phase space $(\phi_e, \pi_e)$ which is the position of the transition from the USR stage to the SR stage. 

In using Eq. (\ref{N-USR}) a few points should be considered. First, the variables $(\phi, \pi)$ represents the initial value in the phase space, i.e. $\phi_{\text{in}}$ and $\pi_{\text{in}}$. More specifically, when we perform $\delta N$, we take the variation with respect to the initial values of the fields and its momentum. However, to simplify the notation, we have discarded the subscript $\text{``in"}$ in all our $\delta N$ analysis. Second, since $\delta \phi$ is massless and is frozen on superhorizon scales, then $\pi$ is  exponentially  decaying so we neglect the variation $\delta \pi$. In other words, while $N=N(\phi, \pi)$ but in taking the variation, we only consider the derivative with respect to $\phi$. This is implicit  in Eqs.  (\ref{fNL-eq}) and   (\ref{gNL-tauNL}).  

To find the total number of e-folds $N_f$ till the point of end of inflation where  $\phi=\phi_f$, we have to solve the field equation (\ref{KG-SR}) during the SR phase as well. This is calculated in \cite{Cai:2018dkf}, yielding
\ba
\label{Nf}
N_{\mathrm SR}= \frac{1}{\eta_V} \ln\big[ - 2 \eta_V \pi_e - 6 \sqrt{2 \epsilon_V} M_P
\big] + \text{constant} \, .
\ea
It is understood that the dependence of $N_{\mathrm SR}$ to the initial configuration $(\phi, \pi)$ is hidden in the quantity $\pi_e$ through the constraint (\ref{constraint}).
Also, a constant value which does not depend on $\pi_e$ and hence does not contribute in $\delta N$ is separated. Finally, in order for inflation to ends we require $\eta_V\neq0$. However, we ignore its effects in the final results for bispectrum and trispectrum  in the limit of our interest where  $|h| \gg \eta_V$. 

Having obtained $N_{\text{USR}}$ and $N_{\mathrm SR}$, the total number of e-folds starting  from the initial point $(\phi, \pi)$ deep in the USR phase to the end of inflation $(\phi_f, \pi_f)$ is given by,
\ba
\label{N-tot}
N_{\text{tot}} = N_{\text{USR}} + N_{\mathrm SR}= 
\frac{1}{3} \ln \Big[1+ \frac{3 (\phi - \phi_e)}{\pi}\Big]+ \frac{1}{\eta_V} \ln\Big[ - 2 \eta_V \pi_e - 6 \sqrt{2 \epsilon_V} M_P
\Big] +  \text{constant} \, ,
\ea
with the understanding that $\pi_e$ itself is related to $(\phi, \pi)$ via the constraint (\ref{constraint}). 

With $N_{\text{tot}}$ given in Eq. (\ref{N-tot}), we are ready to calculate 
$f_{NL}, g_{NL}$ and $\tau_{NL}$. Starting with bispectrum, using Eq. (\ref{fNL-eq}), we obtain \cite{Cai:2018dkf, Firouzjahi:2023aum},
\ba
\label{fNL_h}
f_{NL}= \frac{5 h^2}{2 (h- 6)^2}\, ,
\ea
in which we have discarded the subleading term containing $\eta_V \ll |h|$. 
In the limit of extreme sharp transition $h\rightarrow -\infty$, we obtain the expected result \cite{Namjoo:2012aa} $f_{NL}=\frac{5}{2}$. 

Continuing to the case of trispectrum, using Eq. (\ref{gNL-tauNL}) we obtain,
\ba
\label{gNL-h}
g_{NL}= \frac{25 h^3}{3 (h-6)^3} \, ,
\ea
and
\ba
\label{tau-h}
\tau_{NL}=\frac{9 h^4}{ (h-6)^4} = \frac{36}{25} f_{NL}^2 \, .
\ea
In particular, the second formula for $\tau_{NL}$ confirms that the Suyama-Yamaguchi equality \cite{Suyama:2007bg} does hold even in the presence of the sharpness parameter  $h$.  Furthermore, in the limit of an extreme sharp transition $h\rightarrow -\infty$, the above expressions yield,
\ba
\label{sharp}
g_{NL}= \frac{25}{3} \, , \quad \quad
\tau_{NL}= 9 \, \quad \quad (h \rightarrow -\infty) \, .
\ea 

On the other hand, in the limit of a mild transition $|h| \ll 1$, from Eqs. (\ref{gNL-h})
and (\ref{tau-h}) we see that both $g_{NL}$ and $\tau_{NL}$ approach zero. This is inline with the conclusions of \cite{Cai:2018dkf} who showed that 
$f_{NL}$ approaches zero in the limit of a mild transition as well. The reason is that in the limit of a mild transition, it takes a long time for the mode to settle to its attractor value so during this transition from non-attractor to attractor stage, much of bispectrum and trispectrum are washed out. Indeed, Eqs. (\ref{gNL-h})
and (\ref{tau-h}) suggest that the maximum value of $g_{NL}$ and $\tau_{NL}$ are $g_{NL}=\frac{25}{3}$ and $\tau_{NL}=9$ obtained in the limit of an infinite sharp transition in which the system reaches its attractor phase immediately after the transition.

The above results complete our study of $\delta N$ formalism in calculating the bispectrum parameters $g_{NL}$ and $\tau_{NL}$. In the next section we repeat the analysis of trispectrum using the QFT in-in analysis and confirm the above results. 
In this process, one can also judge the immense simplicity of using $\delta N$ formalism compared to in-in formalism.

\section{ Trispectrum from In-In Formalism}
\label{tri-in-in}

In this section we calculate the trispectrum $\big\langle \mathcal{R}_{\bfk_1}(\tau_0)\mathcal{R}_{\bfk_2}(\tau_0)\mathcal{R}_{\bfk_3}(\tau_0)\mathcal{R}_{\bfk_4}(\tau_0)\big\rangle $
and the parameters $g_{NL}$ and $\tau_{NL}$ using QFT in-in formalism. 

Consider the quantum operator $\hat{O}(\tau)$ whose expectation values 
are calculated at the end of inflation $\tau_0$. For example, for our trispectrum analysis, this corresponds to $\hat{O} = \mathcal{R}_{\bfk_1}\mathcal{R}_{\bfk_2}\mathcal{R}_{\bfk_3}\mathcal{R}_{\bfk_4}$. Within in-in formalism, $ \langle \hat O(\tau_0) \rangle$ is given by \cite{Weinberg:2005vy},
 \begin{equation}
 \label{dyson}
 \langle \hat O(\tau_0) \rangle =   \bigg  \langle 0\Big|  \bigg[ \bar {\mathrm{T}} \exp \bigg( i \int_{-\infty}^{\tau_0} d \tau H_{in} (\tau) \bigg) \bigg] \,  \hat O[\tau_0]  \, \bigg[ \mathrm{T} \exp \bigg( -i \int_{-\infty}^{\tau_0} d \tau H_{in} (\tau) \bigg) \bigg]
 \bigg  |0 \Big\rangle \, ,
 \end{equation}
in which $H_{in}$ represents the interaction Hamiltonian while  $T$ and $\bar T$ denote the time ordering and anti time ordering respectively. In performing the analysis, one considers the first few perturbative expansion of the above compact expression. In addition, the mode functions are calculated in the interaction picture, i.e. the mode function satisfies the free Mukhanov-Sasaki equation in the absence of interaction, given by Eqs.  (\ref{onephase}) and (\ref{R-alpha-beta}) in each stage during inflation. 
Finally, $|0\rangle$ represents the vacuum of the free theory in the absence of interaction.

One technical difficulty in employing the in-in formalism to calculate  
trispectrum is that one needs both the cubic and the quartic Hamiltonians. 
 The cubic action to study the bispectrum was calculated in details by Maldacena \cite{Maldacena:2002vr}. However, the quartic  action is far more complicated when calculated via conventional perturbation theory approach,   see \cite{Seery:2006vu, Jarnhus:2007ia, Arroja:2008ga, Arroja:2009pd,   Mizuno:2009cv, Mizuno:2009mv} for earlier works on quartic Hamiltonians. This difficulty was bypassed in  \cite{Firouzjahi:2023aum} in which the  
quartic action is calculated with reasonable ease via 
effective field theory (EFT) formalism of inflation \cite{Cheung:2007st, Cheung:2007sv}. The cubic and quartic actions are employed in \cite{Firouzjahi:2023aum} to calculate the one-loop corrections in power spectrum in the three-stage model of SR-USR-SR inflation studied in \cite{Kristiano:2022maq, Kristiano:2023scm}. On the other hand, to calculate higher order loop corrections, quintic, sextic and higher orders Hamiltonians  are required.  This question was studied systematically in \cite{Firouzjahi:2025gja, Firouzjahi:2025ihn} who presented a non-perturbative expression for the interaction Hamiltonian within the EFT formalism. 

The EFT formalism is based on the dynamics of the Goldstone boson field $\pi(x^\mu)$ which describes the breaking of the four-dimensional space-time reparameterization invariance to a three-dimensional diffeomorphism invariance in an FLRW background. In EFT approach, one allows all interactions which are permitted by the remaining three-dimensional  diffeomorphism invariance.  However, the great advantage of the EFT approach appears when one works in the decoupling limit where the gravitational back-reactions are neglected. Practically, this mean one can neglect the perturbations in the laps and shift functions in ADM formalism and considers only the perturbations from the matter field sector.  

Employing the EFT approach, the cubic and quartic Hamiltonian in USR-SR setup  
are given by \cite{Firouzjahi:2023aum}, 
\begin{equation}
\label{H3ham}
{\bf H}_3=-H^3\eta\epsilon a^2\int d^3x \Big(\pi\pi'^2-\pi(\partial \pi)^2 \Big) \, ,
\end{equation}
and, 
\begin{equation}
\label{H4ham}
{\bf H}_4=\frac{1}{2}\int d^3x \Big[(H^4\eta^2\epsilon a^2-\eta'H^3\epsilon a)\pi^2\pi'^2+(H^4\eta^2\epsilon a^2+\eta'H^3\epsilon a)\pi^2(\partial\pi)^2 \Big],
\end{equation}
where a prime in this and next section means the derivative with respect to the conformal time $d \tau= dt/a(t)$. Note that $\epsilon(\tau)$ and $\eta(\tau)$ should be calculated during each stage of inflation with the appropriate values as given in Section \ref{model}. In particular, note that $\eta' \propto \delta (\tau-\tau_e)$ as given in Eq. (\ref{eta-jump}). 

The above interaction Hamiltonians are written in terms of the Goldstone field $\pi(x^\mu)$. However, we are interested in trispectrum associated to $\calR$. Therefore, we need a non-linear dictionary between $\pi$ and $\calR$. To the cubic orders in perturbations which are required for the trispectrum analysis, this dictionary is given by \cite{Firouzjahi:2023aum}, 
 \begin{eqnarray}
 \label{nonlinear}
 \calR &=& - H \pi + \big( H \pi \dot \pi + \frac{\dot H}{2} \pi^2 \big)
 + \big( -H \pi \dot \pi^2 -\frac{H}{2} \ddot \pi \pi^2 - \dot H \dot \pi \pi^2 -\frac{\ddot H}{6} \pi^3 \big)  + {\cal O}( \pi^4) \, .
 \end{eqnarray}

We calculate the trispectrum at the end of inflation when the system has reached its attractor phase  during which $H$ and $\pi$ are nearly constants. In this limit, 
all higher orders corrections in Eq. (\ref{nonlinear}) are subleading and one is left with the linear relation $ \mathcal{R}=-H\pi+\mathcal{O}(\epsilon\pi^2)$. Therefore, 
\ba
 \langle \calR_{\bfk_1}  \calR_{\bfk_2} \calR_{\bfk_3}  \calR_{\bfk_4}\rangle
= H^4 \langle \pi_{\bfk_1}  \pi_{\bfk_2} \pi_{\bfk_3}  \pi_{\bfk_4}  \rangle   \, ,
\ea
so the trispectrum of $\calR$ is proportional to the trispectrum of $\pi$. 

\begin{figure}
  \center
   \includegraphics[width=0.5\textwidth]{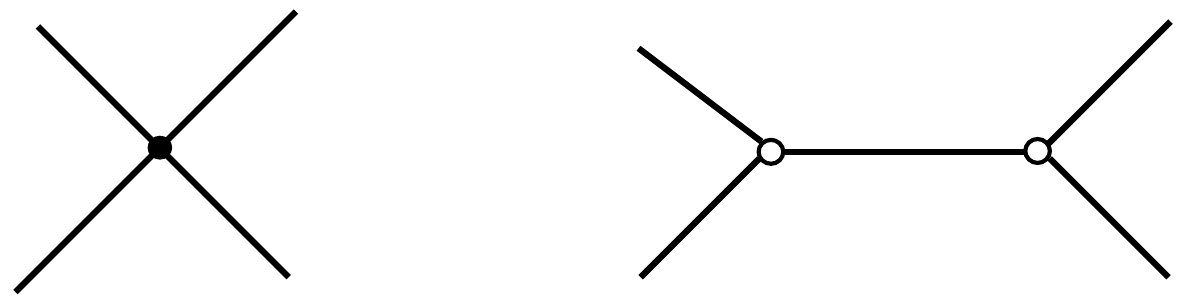}
   \vspace{.7cm}
       \caption{The Feynman diagrams associated to the trispectrum. The 
       filled (empty) circle  represents the quartic (cubic) Hamiltonians. The left (right) diagram represents the contribution of ${\bf H}_4$  (${\bf H}_3$ ).  }
\label{diagrams}
\vspace{0.7cm}
\end{figure}

There are two different types of Feynman diagrams relevant for the trispectrum as presented in Fig. \ref{diagrams}. The left panel involves a single vertex of ${\bf H}_4$ while the right panel contains two vertices of ${\bf H}_3$.  The contribution of the left diagram is easier to handle since it involves a single time integral. However, the analysis of right diagram is far more complicated since it involves two vertices of  ${\bf H}_3$, meaning it involves  a double nested time integrals from the expansion of the master formula (\ref{dyson}).  Below we calculate the contributions of each diagram starting with the left diagram.

\subsection{Contributions from ${\bf H}_4$}
\label{H4-result}

Here we present the contributions from the left diagram in Fig. \ref{diagrams} involving a single vertex of quartic Hamiltonian ${\bf H}_4$.
Expanding the Dyson series in master formula (\ref{dyson}) to first order in ${\bf H}_4$ yields, 
\ba
\label{H4-a}
\big\langle \mathcal{R}_{\bfk_1}(\tau_0)\mathcal{R}_{\bfk_2}(\tau_0)\mathcal{R}_{\bfk_3}(\tau_0)\mathcal{R}_{\bfk_4}(\tau_0)\big\rangle = - 2 \mathrm{Im} \int_{-\infty}^{\tau_0} \Big\langle {\bf H}_4(\tau) \mathcal{R}_{\bfk_1}(\tau_0)\mathcal{R}_{\bfk_2}(\tau_0)\mathcal{R}_{\bfk_3}(\tau_0)\mathcal{R}_{\bfk_4}(\tau_0) \Big\rangle\, ,
\ea
in which ${\bf H}_4$ is given in Eq. (\ref{H4ham}). As we see, ${\bf H}_4$ has two different terms, the time derivative term and the gradient term. As can be checked easily, the contributions from the gradient term are sub-leading as they yield the factors like $k \tau_e$ which are suppressed in the superhorizon limit. The reason is that we calculate the trispectrum for the modes which leaves the horizon during the USR phase so they are superhorizon by the time of the transition. Of course, if one is interested, one can keep these sub-leading terms in the situations where  the modes are not quite superhorizon at  $\tau_e$. 

Discarding the contributions of the gradient term, we obtain,
\ba
\label{H4-b}
\big\langle \mathcal{R}_{\bfk_1}(\tau_0)\mathcal{R}_{\bfk_2}(\tau_0)\mathcal{R}_{\bfk_3}(\tau_0)\mathcal{R}_{\bfk_4}(\tau_0)\big\rangle = 
\frac{-M_P^2}{H^4} \,  \mathrm{Im} \int_{-\infty}^{\tau_0} f(\tau)
\Big\langle  \calR(\tau)^2 \calR'(\tau)^2 \mathcal{R}_{\bfk_1}\mathcal{R}_{\bfk_2}\mathcal{R}_{\bfk_3}\mathcal{R}_{\bfk_4}(\tau_0)
\Big\rangle \, ,
\ea
in which the function $f(\tau)$ is given by,
\ba
\label{f-def}
f(\tau) \equiv H^4\eta^2\epsilon a^2-\eta'H^3\epsilon a \, .
\ea
In particular, note that $f(\tau)$ contains a local term $\delta(\tau-\tau_e)$ from
$\eta'$. 

The domain of the time integral in Eq. (\ref{H4-b}) is $(-\infty, \tau_0)$ so we have three different contributions,  (1): the first contribution is from the bulk of the USR, corresponding to region $(-\infty, \tau_e^-)$. (2): the second contribution is from the local source $\eta'$ which involves the $\delta(\tau-\tau_e)$ term. Finally, (3): the third contribution is from the SR region  $(\tau_e^+, \tau_0)$.
  Note that for each region of integration we should use the corresponding values of $\epsilon(\tau), \eta(\tau)$ and the mode function $\calR(\tau)$. Below we present each contribution in turn labeled by their orders, (1), (2) and (3) as listed above. 

Performing the in-in integral, from the bulk of the USR region, we obtain,
\begin{equation}
\label{reg1}
\big\langle \mathcal{R}_{\bfk_1}\mathcal{R}_{\bfk_2}\mathcal{R}_{\bfk_3}\mathcal{R}_{\bfk_4}(\tau_0)\big\rangle_{(1)}' =
\frac{9  (h-6)^3 (h+18)}{32  h^4}
\frac{H^6}{M_P^6 \epsilon_e^3}
\frac{\sum_i k_i^3}{\prod_i k_i^3}\, \quad \quad
(\text{bulk of USR})
\end{equation}
in which $\langle \rangle'$ means that we have absorbed the overall factor 
$(2 \pi)^3 \delta^3(\sum_i \bfk_i)$. 

From the local source induced by $\eta'$, we obtain,
\begin{equation}
\label{reg2}
\big\langle \mathcal{R}_{\bfk_1}\mathcal{R}_{\bfk_2}\mathcal{R}_{\bfk_3}\mathcal{R}_{\bfk_4}(\tau_0)\big\rangle_{(2)}' =
\frac{9  (h-6)^3 (h+6)}{64  h^3}
\frac{H^6}{M_P^6 \epsilon_e^3}
\frac{\sum_i k_i^3}{\prod_i k_i^3}\, \quad \quad
(\text{local source } \eta')
\end{equation}

Finally, from the SR region $(\tau_e^+, \tau_0)$, we obtain,
\begin{equation}
\label{reg2}
\big\langle \mathcal{R}_{\bfk_1}\mathcal{R}_{\bfk_2}\mathcal{R}_{\bfk_3}\mathcal{R}_{\bfk_4}(\tau_0)\big\rangle_{(3)}' =
-\frac{9  (h-6)^3 (h+6)}{640  h^6}   \big( 10 h^3 + 45 h^2 + 72 h- 108 \big)  
\frac{H^6}{M_P^6 \epsilon_e^3}
\frac{\sum_i k_i^3}{\prod_i k_i^3}\, \quad \quad
(\text{SR }) 
\end{equation}

Adding the above three terms, the total contribution from ${\bf H}_4$ is given by,
\begin{equation}
\label{reg2}
\big\langle \mathcal{R}_{\bfk_1}\mathcal{R}_{\bfk_2}\mathcal{R}_{\bfk_3}\mathcal{R}_{\bfk_4}(\tau_0)\big\rangle_{{\bf H}_4}' =
-\frac{9  (h-6)^3 (h+6)}{640  h^6}   \big( 25 h^3 -18 h^2 + 324 h- 648 \big)  
\frac{H^6}{M_P^6 \epsilon_e^3}
\frac{\sum_i k_i^3}{\prod_i k_i^3}\, . 
\end{equation}

Noting that $P_\calR(k) \propto k^{-3}$, the above expression indicates that the trispectrum induced by the quartic Hamiltonian contributes only into $g_{NL}$ and has no contribution in $\tau_{NL}$. More specifically, using Eq. (\ref{PR-USR}) for 
$P_\calR(\tau_0, k)$ and the definition of trispectrum in Eq. (\ref{TR}), we obtain,
\ba
\label{TR-H4}
T_\calR(k_1, k_2, k_3, k_4)\big|_{{\bf H}4}= 
\frac{ -9(25 h^3 -18 h^2 + 324 h- 648)}{10 (h-6)^3}
 \Big[ P_\calR(k_{2}, \tau_0) P_\calR(k_3, \tau_0) P_\calR(k_4, \tau_0)
+ 3\,  \text{perms.} \Big]. \nonumber\\
\ea

\subsection{Contributions from ${\bf H}_3$}
\label{H3TR}

Here we present the contributions from the diagram in the right panel  of Fig. \ref{diagrams}. As mentioned before, the analysis associated to this diagram  is far more complicated compared to the left diagram since we have to perform a double nested time integrals. We relegate the details of the analysis into Appendix \ref{H3} and here present the main results.

Expanding the in-in integrals in Dyson series (\ref{dyson}) to second order in ${\bf H}_3$ yields, 
\ba
   \langle \hat O(\tau_0)  \rangle_{{\bf H}3} =    \langle \hat O(\tau_0) \rangle_{(2,0)} +   \langle \hat O(\tau_0) \rangle_{(1,1)} +    \langle \hat O(\tau_0) \rangle_{(0, 2)}
  \ea
in which
\ba
 \label{20-int}
  \langle \hat O(\tau_0) \rangle_{(2,0)} 
=
- \int_{-\infty}^{\tau_0} d \tau_1 \int_{-\infty}^{\tau_1} d \tau_2
\big \langle {\bf H}_3 (\tau_2)  {\bf H}_3 (\tau_1) \hat O(\tau_0)
 \big \rangle  
= \langle \hat O(\tau_0)
\rangle^\dagger_{(0,2)}\, ,
\ea
and
\begin{equation}
  \label{11-int}
\langle \hat O(\tau_0) \rangle_{(1,1)} =
 \int_{-\infty}^{\tau_0} d \tau_1 \int_{-\infty}^{\tau_0} d \tau_2
\big \langle {\bf H}_3 (\tau_1)   \hat O(\tau_0)
{\bf H}_3 (\tau_2) \big \rangle   \, ,
\end{equation}
in which in our case $\hat O(\tau_0) =  \mathcal{R}_{\bfk_1}\mathcal{R}_{\bfk_2}\mathcal{R}_{\bfk_3}\mathcal{R}_{\bfk_4}(\tau_0)$. 

In calculating the above integrals, as discussed before,  we discard the contribution in ${\bf H}_3$ which involve the gradient terms as they lead to subleading contributions in trispectrum on superhorizon limit. Depending on whether $\tau_1$ and $\tau_2$ are in the USR region or SR region and noting that because of time-ordering $\tau_2 \leq \tau_1$, there are  three possibilities for the range of the double integrals over $(\tau_1, \tau_2)$ as   (USR-USR), (SR-USR) and (SR-SR). We present the results for each case separately while for further details see  Appendix \ref{H3}.

\subsubsection{ USR-USR}

First consider the case when both $\tau_1$ and $\tau_2$ are in USR region. In this case, we obtain,
\ba
\label{USR-USR}
T_\calR(k_1, k_2, k_3, k_4)\big|_{USR-USR} &=&\Big( \frac{H^6}{M_P^6 \epsilon_e^3} \Big) 
\Big[ -\frac{81  (h-6)^3 }{16  h^4}\Big(\frac{\sum_i k_i^3}{\Pi_i k_i^3} \Big)
\nonumber\\&+&\frac{9  (h-6)^2 (h+12)^2 }{64 h^4} \Big(\frac{1}{k_1^3 k_2^3 k_{13}^3}+\text{11 perms.}  \Big) \Big] ,
\ea
where $k_{ij}\equiv | \bfk_i+\bfk_j |$. We see that the above result for trispectrum contributes into both $g_{NL}$ 
and $\tau_{NL}$.

\subsubsection{ SR-USR}

Now consider the case when $\tau_1$ is in SR region while $\tau_2$ is in USR region. In this case, we obtain, 
\ba
\label{SR-USR}
T_\calR(k_1, k_2, k_3, k_4)\big|_{SR-USR} &=&\Big( \frac{H^6}{M_P^6 \epsilon_e^3} \Big)  \Big[
\frac{27  (h-6)^3 (h+6) }{64 h^4} \Big(\frac{\sum_i k_i^3}{\Pi_i k_i^3} \Big) \nonumber\\
&-&\frac{9 
 (h-6)^2 (h+6) (h+12) }{16 h^4} \Big(\frac{1}{k_1^3 k_2^3 k_{13}^3}+\text{11 perms.} \Big) \Big] \, .
\ea

\subsubsection{ SR-SR}
Finally, consider the case when both $\tau_1$ and $\tau_2$ are in the USR region. This case yields to, 
\ba
\label{SR-SR}
T_\calR(k_1, k_2, k_3, k_4)\big|_{SR-SR} &=&\Big( \frac{H^6}{M_P^6 \epsilon_e^3} \Big)  \Big[
\frac{27  (h-6)^3 (h+6)^2 (5 h-6) }{640 h^6} \Big(\frac{\sum_i k_i^3}{\Pi_i k_i^3} \Big)\nonumber\\
&+&\frac{9 \left(h^2-36\right)^2 }{16 h^4} \Big(\frac{1}{k_1^3 k_2^3 k_{13}^3}+\text{11 perms.}\Big) \Big] \, .
\ea

Adding the above three results (\ref{USR-USR}), (\ref{SR-USR}) and (\ref{SR-SR}), the total contribution of the right diagram in Fig. \ref{diagrams}  is obtained to be,
\ba
\label{TR-H3a}
T_\calR(k_1, k_2, k_3, k_4)\big|_{{\bf H}_3} =\Big( \frac{H^6}{M_P^6 \epsilon_e^3} \Big)  \Bigg[ \frac{81  (h-6)^3}{640 h^6}  ( 5 h^3
 - 2 h^2 + 36 h- 72) \frac{\sum_i k_i^3}{\prod_i k_i^3}
 \nonumber\\
\quad  +  \frac{9  (h-6)^2 }{64 h^2} \Big(\frac{1}{k_1^3 k_2^3 k_{13}^3}+\text{11 perms} \Big) \Bigg] \, .
\ea

Expressing the above results in terms of $P_\calR(k, \tau_0)$, we finally obtain,
\ba
\label{TR-H3}
T_\calR(k_1, k_2, k_3, k_4)\big|_{{\bf H}_3} &=& 
\frac{9h^4}{(h-6)^4} 
 \Big[ P_\calR(k_{13}) P_\calR(k_3) P_\calR(k_4) + (11~ \text{perms}) \Big] \\
&+&  \frac{81}{10 (h-6)^3} ( 5 h^3
 - 2 h^2 + 36 h- 72)
\Big[ P_\calR(k_{2}) P_\calR(k_3) P_\calR(k_4)
+ (3~ \text{perms}) \Big]\, . \nonumber
\ea

\subsection{Total Trispectrum}
\label{total}

Having the contributions of each Feynman diagram as given in Eqs. (\ref{TR-H4}) and  (\ref{TR-H3}), the total trispectrum is obtained to be,
\ba
\label{TR-tot}
T_\calR(k_1, k_2, k_3, k_4)\big|_{\text{tot}} &=&  \frac{9h^4}{(h-6)^4}
 \Big[ P_\calR(k_{13}, \tau_0) P_\calR(k_3, \tau_0) P_\calR(k_4, \tau_0) + (11~ \text{perms}) \Big] \nonumber\\
&+&   \frac{18 h^3}{(h-6)^3}\Big[ P_\calR(k_{2}, \tau_0) P_\calR(k_3, \tau_0) P_\calR(k_4, \tau_0)
+ (3~ \text{perms}) \Big] \, .
\ea
As expected, the trispectrum has the local shape defined in Eq. (\ref{TR}).
Furthermore,  calculating  the coefficients $g_{NL}$ and $\tau_{NL}$, the results agree exactly with  Eqs. (\ref{gNL-h}) and (\ref{tau-h}) obtained via 
$\delta N$ formalism.  It is interesting and reassuring that  the trispectrum obtained from $\delta N$ formalism and in-in approach match with each other exactly.   This also confirms the validity of EFT formalism in constructing the interaction Hamiltonian and also the validity of the decoupling limit  employed in our EFT approach.

\section{Trispectrum in Setup with Infinitely Sharp Transition }
\label{sharp-tran}

In the above in-in analysis, we have considered a general value of 
the sharpness parameter  $|h| >1$ and have calculated the trispectrum at the time of end of inflation $\tau_0$ when the system has reached its attractor phase.
However, a particular case is when the transition from the USR phase to the SR phase is infinitely sharp, $h \rightarrow -\infty$. In this case, the system reaches the attractor phase immediately after the transition, say at $\tau_e^+$. In this particular case, one does not need to wait till the time of end of inflation to calculate the cosmological correlations such as the bispectrum or trispectrum. Instead, one can calculate them right at $\tau=\tau_e$. Indeed, this was the method which was employed to calculate the bispectrum 
in the original works \cite{Namjoo:2012aa, Chen:2013aj} since in these works $h\rightarrow -\infty$. 
This will bring simplification as one does not need to go through the analysis in the (SR, USR) and (SR-SR) stages outlined in previous section.
However, a new technical difficulty occurs that one has to take into account the non-linear relation between $\pi$ and $\calR$ given in Eq. (\ref{nonlinear}). 
This is an interesting exercise which highlights the non-linear relation between $\calR$ and $\pi$ which worth considering.

The contribution of ${\bf H}_3$ will be just the contribution from the (USR-USR) region which was calculated already in Eq. (\ref{USR-USR}). Now setting $h \rightarrow -\infty$, we obtain, 
\ba
\label{USR-USR2}
T_\calR(k_1, k_2, k_3, k_4;\tau_e)\big|_{USR-USR} 
=
\frac{9   }{64} \Big(\frac{1}{k_1^3 k_2^3 k_{13}^3}+\text{11 perms.}  \Big) \, .
\ea

As for ${\bf H}_4$, we have only two contributions. The first contribution is from the bulk which is the same as in Eq. (\ref{reg1}) with $h\rightarrow -\infty$. This yields,
\begin{equation}
\label{reg1-b}
T_\calR(k_1, k_2, k_3, k_4; \tau_e)\big|_{(1)}  =
\frac{9  }{32 }
\frac{H^6}{M_P^6 \epsilon_e^3}
\frac{\sum_i k_i^3}{\prod_i k_i^3}\, \quad \quad
(\text{bulk of USR}) \, .
\end{equation}

The second contribution is from the local source term involving $\eta' \propto \delta(\tau- \tau_e)$. The important point here is that the integral is cut right at $\tau_e$ i.e. we should consider the domain $(\tau_e^-, \tau_e)$ instead of the region   $(\tau_e^-, \tau_e^+)$.  To do this properly, we should consider $\eta(\tau)$ as \cite{Namjoo:2012aa} 
$\eta(\tau)= -6(1-\theta(\tau-\tau_e))$. Furthermore, as we consider only the interval $(\tau_e^-, \tau_e^+)$,  this brings a factor $\frac{1}{2}$ because,
\ba
\int_{-\infty}^0 dx \,  \delta(x) =\frac{1}{2} \, .
\ea
With the above prescription in mind, and performing the in-in integral, we obtain,  
\begin{equation}
\label{reg2-b}
T_\calR(k_1, k_2, k_3, k_4;\tau_e)\big|_{(2)} =
-\frac{27  }{64}
\frac{H^6}{M_P^6 \epsilon_e^3}
\frac{\sum_i k_i^3}{\prod_i k_i^3}\, \quad \quad
(\text{local source } \eta') \, .
\end{equation}

Now the new contributions we are dealing with are from the non-linear relation between $\calR$ and $\pi$ in   Eq. (\ref{nonlinear}). Dropping the time derivatives of $H$ which are slow-roll suppressed,  we obtain,
\begin{eqnarray}
 \label{nonlinear2}
 \calR &\simeq& - H \pi +  H \pi \dot \pi 
 + \big( -H \pi \dot \pi^2 -\frac{H}{2} \ddot \pi \pi^2  \big)  + {\cal O}( \pi^4) \, .
 \end{eqnarray}  
Let us start with the quadratic correction, $H \pi \dot \pi$. Its contribution in trispectrum has the following form,
\ba
\label{first}
\big \langle \calR_{\bfk_1} \calR_{\bfk_2} \calR_{\bfk_3} \calR_{\bfk_4}
\big \rangle &\rightarrow&  \big \langle ( H \pi \dot \pi)_{\bfk_1} \calR_{\bfk_2} \calR_{\bfk_3} \calR_{\bfk_4} \big \rangle + 3 ~  \text{perms.}  \nonumber\\
&+& 
 \big \langle ( H \pi \dot \pi)_{\bfk_1} ( H \pi \dot \pi)_{\bfk_2} \calR_{\bfk_3} \calR_{\bfk_4} \big \rangle + 6 ~  \text{perms.}
\ea  
However, to calculate the correlation from the first line above  we have to use the in-in formalism once more since the correlator involves an odd number of the Gaussian fields, see Appendix \ref{NLterms} for more details. 
Specifically, 
\ba
\big \langle (  \pi \dot \pi)_{\bfk_1} \calR_{\bfk_2} \calR_{\bfk_3} 
\calR_{\bfk_4}(\tau_e)
\big \rangle= - 2 \mathrm{Im} \int \frac{d^3 q}{(2 \pi)^3} \int_{-\infty}^{\tau_e}
d \tau \Big \langle {\bf H}_3 (\tau) \pi_{\bfq} \pi'_{\bfk_1- \bfq} (\tau_e)\calR_{\bfk_2} \calR_{\bfk_3} \calR_{\bfk_4}(\tau_e)
\Big \rangle \, .
\ea
Note that all operators next to ${\bf H}_3 (\tau) $ are at the time $\tau_e$. Also note that the integral $d^3 q$ comes from the convolution integral in Fourier space. 

Calculating the above integral, the contribution of the quadratic part $H \pi \dot \pi$ from the first line of Eq. (\ref{first})  yields,
 \begin{equation}
\label{PiR2a} 
T_\calR(k_1, k_2, k_3, k_4;\tau_e)\big|_{(H \pi \dot \pi)_1} =  
\Big(\frac{H^6}{M_P^6 \epsilon_e^3} \Big) 
\Big[ -\frac{27 }{32  }\frac{\sum_i k_i^3}{\prod_i k_i^3}\\-\frac{9  }{16 } \big(\frac{1}{k_1^3k_2^3k_{13}^3}+11\text{perms.} \big) \Big] \, .
 \end{equation}

On the other hand, calculating the second line in Eq. (\ref{first}) is easy 
since it involves the correlations of 
even number of Gaussian fields and it does not require the in-in integral, yielding 
\begin{equation}
\label{PiR2b}
T_\calR(k_1, k_2, k_3, k_4;\tau_e)\big|_{(H \pi \dot \pi)_2} =  
\frac{9  }{16 } \Big(\frac{H^6}{M_P^6 \epsilon_e^3} \Big) 
  \big(\frac{1}{k_1^3k_2^3k_{13}^3}+11\text{perms.} \big) \, .
\end{equation} 

Finally, we have to include the contributions of the cubic corrections between $\calR$ and $\pi$, the correction from the  remaining two terms in Eq. (\ref{nonlinear2}). As in the  second line of Eq. (\ref{first}), they lead to the correlations of even number of fields  so  they do not need in-in integrals, yielding, 
\begin{equation}
\label{PiR3a}
T_\calR(k_1, k_2, k_3, k_4;\tau_e)\big|_{(\pi \dot\pi^2)} =  
\frac{27  }{32 } \Big(\frac{H^6}{M_P^6 \epsilon_e^3} \Big) 
 \frac{\sum_i k_i^3}{\prod_i k_i^3} \, ,
\end{equation}
and
\begin{equation}
\label{PiR3b}
T_\calR(k_1, k_2, k_3, k_4;\tau_e)\big|_{(\pi^2 \ddot\pi)} =  
\frac{27  }{64 } \Big(\frac{H^6}{M_P^6 \epsilon_e^3} \Big) 
 \frac{\sum_i k_i^3}{\prod_i k_i^3} \, .
\end{equation}

Adding all contributions from Eqs. (\ref{USR-USR2}), (\ref{reg1-b}), (\ref{reg2-b}),  (\ref{PiR2a}),  (\ref{PiR2b}), (\ref{PiR3a}) and (\ref{PiR3b}) and using Eq. (\ref{PR-USR}) to  express the final results in terms of power spectrum $P_\calR(k_{i}, \tau_0)$, we finally obtain, 
\ba
\label{TR-infinite}
T_\calR(k_1, k_2, k_3, k_4)\big|_{\text{tot}} &=&  9
 \Big[ P_\calR(k_{13}, \tau_0) P_\calR(k_3, \tau_0) P_\calR(k_4, \tau_0) + (11~ \text{perms}) \Big] \nonumber\\
&+&   18\Big[ P_\calR(k_{2}, \tau_0) P_\calR(k_3, \tau_0) P_\calR(k_4, \tau_0)
+ (3~ \text{perms}) \Big] \, .
\ea
This is in agreement with the full result Eq. (\ref{TR-tot}) in the limit 
$h \rightarrow -\infty$. Furthermore, in this limit we obtain $g_{NL}=\frac{25}{3}$ and $ \tau_{NL}=9$ as obtained via $\delta N$ formalism 
in Eq. (\ref{sharp}). 

The morale of this analysis was to demonstrate the effects of 
non-liner relation between $\pi$ and $\calR$. In the case when $h \rightarrow -\infty$, the system reaches the attractor phase immediately after the transition so one can calculate the trispectrum right at $\tau=\tau_e$. However, one has to take the non-linear relation between $\pi$ and $\calR$ into account.  Compare this to our analysis in section \ref{tri-in-in} where we have calculated the correlations at the time of end of inflation when the mode function is frozen  and the non-linear relation between $\pi$ and $\calR$ was not important.

\section{Shape of Trispectrum}

\begin{figure}
\center
\vspace{-0.5cm}
  \includegraphics[width=0.41\textwidth]{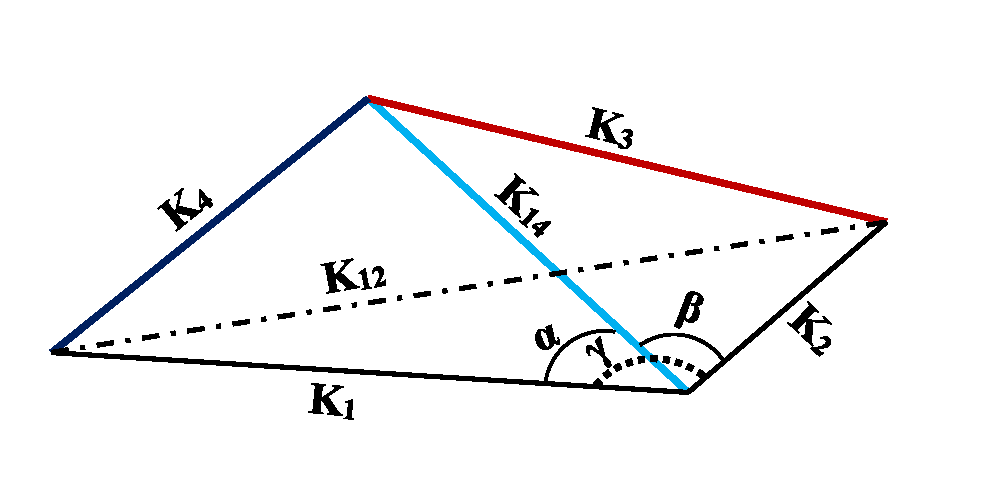}\caption{A schematic view of the tetrahedron constructed by  $\bfk_i$.}
 \label{geom} 
\end{figure}

In this section we briefly study the shape of trispectrum obtained in 
Eq. (\ref{TR-tot}) in the local limit. A schematic view of the momenta $\bfk_i$ are presented in Fig. \ref{geom}. The trispectrum is a function of six independent momenta, $k_1, k_2, k_3, k_4, k_{12}, k_{14}$. 

Following the convention of \cite{Chen:2009bc}, let us define the angles $\alpha, \beta$ and $\gamma $ as follows, 
\begin{align}
\label{geomeq}
  \cos(\alpha) &= \frac{k_1^2+k_{14}^2-k_4^2}{2k_1k_{14}}~, \nonumber \\
\cos(\beta) &= \frac{k_2^2+k_{14}^2-k_3^2}{2k_2k_{14}}~, \nonumber \\
\cos(\gamma) &= \frac{k_1^2+k_{2}^2-k_{12}^2}{2k_1k_{2}}~.
\end{align}
Then in order for $\bfk_i$ to form a tetrahedron, the condition 
$\cos(\alpha-\beta) \ge \cos(\gamma) \ge \cos(\alpha+\beta)$ should be satisfied.   Furthermore, all triangle inequalities such as $k_1+ k_4 > k_{14}, k_1+ k_2 > k_{12}$ etc should be satisfied as well.  

\begin{figure}[h!]
\center
  \includegraphics[width=0.41\textwidth]{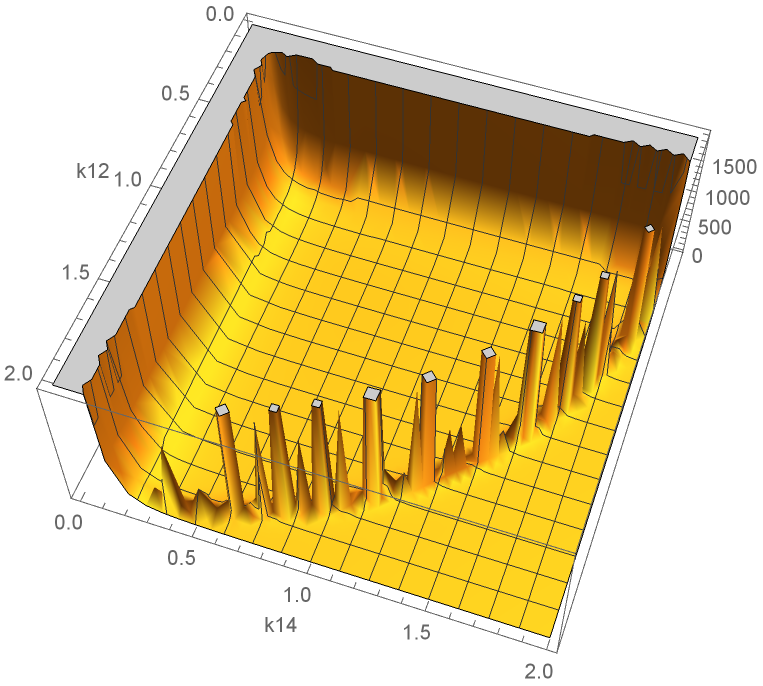}
 \vspace{0.5cm}
  \caption{The non-planar equilateral shape of trispectrum in which  $k_1=k_2=k_3=k_4$ with $h=-6$.
  The peaks are along the axis $k_{12}, k_{14} \rightarrow 0$ and along  the circle determined by  $k_{12}^2+k_{14}^2=4 $. }
  \vspace{0.5cm}
  \label{equil}
\end{figure}

There are various possibilities for the shapes, so for brevity, 
here we present some sample cases for comparison. 
To present the plots of the shape we  can have only two independent variables. Without loss of generality we set $\bfk_1= \hat{\bf i}$. Then, one possible choice is the equilateral configuration in which $k_1=k_2=k_3=k_4$. The shape of this specific case is presented in Fig \ref{equil}. As we see there are  peaks in the figure  with $k_{14}=0$ or $k_{12}=0$  as is expected in local shape. These peaks are located in either boundaries (i.e. $x$ axis and $y$ axis in Fig \ref{equil}). The other case with pronounced peak is the  limit when $k_{13}=0$. However, since 
\ba
k_{13}= \sqrt{k_1^2 + k_2^2 + k_3^2 + k_4^2- k_{12}^2 - k_{14}^2} \, ,
\ea
the condition $k_{13}=0$ corresponds to the circle $k_{12}^2+k_{14}^2=4 $. The peaks associated to this case are clearly seen in  Fig \ref{equil} as well.

\begin{figure}[h!]
\center
  \includegraphics[width=0.45\textwidth]{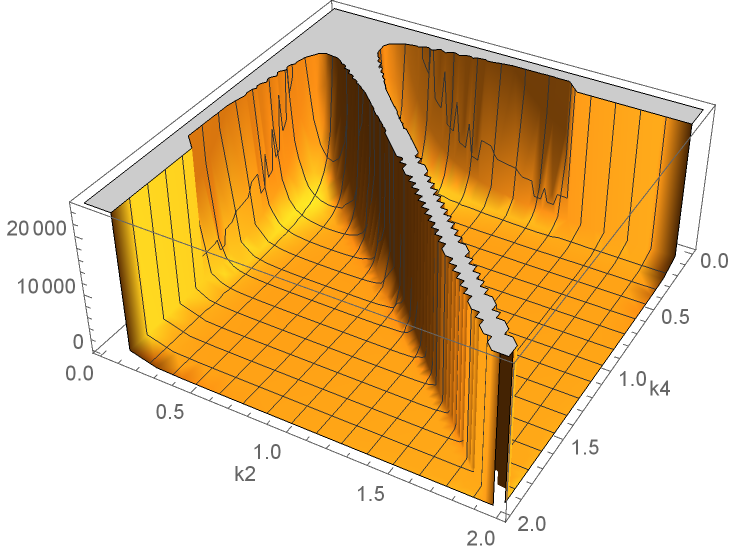}
  \hspace{1cm}
 \includegraphics[width=0.37\textwidth]{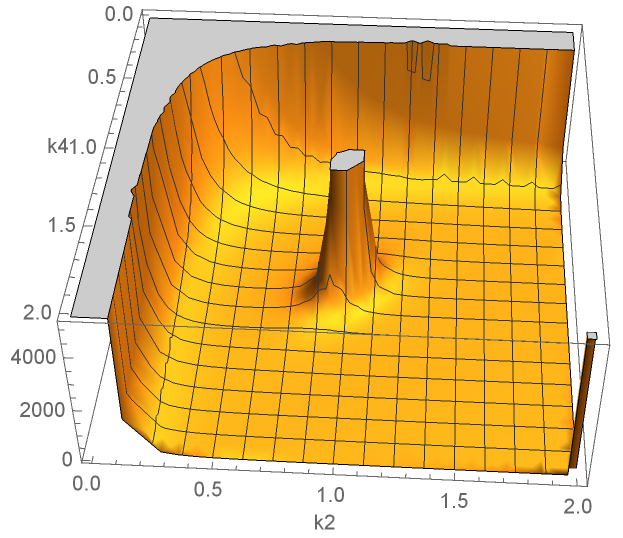}
 \vspace{0.5cm}
 \caption{Left: the case with $\frac{-\pi}{2}\leq \alpha-\gamma\leq 0$ with  the peaks along the axis where $k_2,k_4\rightarrow 0$ and along the line 
where $k_2=k_4$. Right: The case with $0\leq \alpha-\gamma\leq\frac{\pi}{2}$ with the  peaks at  the point $k_2=k_4=1$ and along the axis  $k_2,k_4\rightarrow 0$. In both cases, $h=-6$.  }
 \label{planar}
\end{figure}

Another possible configuration that we can study is the limit where all vectors are in the same plane. One possible choice of vectors is to set $k_1=k_{14}=k_3$ and again without loss of generality we take ${\bfk}_1=\hat {\bf{i}}$. In this case, according to Fig. \ref{geom},  $k_2=2 \cos(\alpha-\gamma)$ and then $\frac{-\pi}{2}\leq \alpha-\gamma\leq \frac{\pi}{2}$. We have presented the shape functions for $\frac{-\pi}{2}\leq \alpha-\gamma\leq 0$ and $0\leq \alpha-\gamma\leq \frac{\pi}{2}$ separately in Fig \ref{planar}. As we see from both panels of this figure, the peaks appear in the boundaries $k_2\rightarrow0$ and $k_4\rightarrow0$ as expected in local shapes. Moreover, for  $\frac{-\pi}{2}\leq \alpha-\gamma\leq 0$ $(\alpha-\gamma=-\text{arccos}(\frac{k_2}{2}))$, the other limit that the shape blows up is when $k_{13}=0$ corresponding to $k_2=k_4$  as seen in the left panel of Fig. \ref{geom}. Finally,  for the  case  $0\leq \alpha-\gamma\leq \frac{\pi}{2}$ $(\alpha-\gamma=\text{arccos}(\frac{k_2}{2}))$  one can show that $k_{12}=0$ corresponds to 
 $k_2=k_4=1$ and again the shape will have a peak in this limit as can be seen in the right panel of Fig. \ref{geom}.

\section{Summary and Discussions}

In this work we have studied the trispectrum in the two-phase 
USR-SR model of inflation.  The USR phase is extended in the past while to keep the setup under perturbative control, the USR phase is terminated by the attractor SR stage followed by reheating as in conventional SR setups. 
An important role is played by the sharpness parameter $h$ which controls how quickly the system reaches the final attractor phase after the USR stage. To simplify the analysis, we have considered the limit of sharp transition $|h|>1$ in which the subleading SR corrections may be ignored. We have assumed that all four modes have left the horizon during the USR stage where the CMB scale perturbations are generated.

We have employed both $\delta N$ and in-in formalisms to calculate the bispectrum.  Each method has its own advantages. $\delta N$ approach has the major advantage that it is simple and more direct. One only needs to follow the background trajectory in phase space. Since the system is in a  non-attractor phase during the USR stage, both the field and its momentum should be considered in this $\delta N$ analysis. In order to employ the $\delta N$ formalism, one needs to assume that all four modes are in superhorizon limit. This brings a limitation in the case  where there are hierarchies between the scales of the modes such that some modes leave the horizon during the follow up SR  phase. On the other hand, the in-in formalism has the advantage that it is based on first principle QFT analysis and does not rely on the assumption of superhorizon limit to be 
employed. However, the analysis proved to be more challenging and long within the in-in formalism. 
In order to calculate the trispectrum via in-in formalism, one needs the cubic and the quartic Hamiltonians. We have employed the formalism of EFT of inflation to calculate the cubic and quartic Hamiltonians in the decoupling limit. We have confirmed that both $\delta N$ and in-in approaches yield the same result for the bispectrum. This also confirms various assumptions imposed during the analysis such as the forms of the interaction Hamiltonians or the validity of the  decoupling limit in EFT formalism.

The trispectrum with general value of the sharpness parameter $h$  is given in Eq. (\ref{TR-tot}) yielding to parameters $g_{NL}$ and $\tau_{NL}$ given in Eqs.  (\ref{gNL-h}) and (\ref{tau-h}). We have shown that the Suyama-Yamaguchi inequality $\tau_{NL}\ge \frac{36}{25} f_{NL}^2$ is saturated with the equality sign as expected for the single field models. A particular case of  interest is the limit of infinitely  sharp transition 
$h \rightarrow -\infty$ in which the system reaches the attractor phase immediately after the USR phase. In this limit, one can calculate the cosmological correlators such as the bispectrum or trispectrum at the time $\tau=\tau_e$. We have studied this case separately, calculating the  trispectrum  at $\tau=\tau_e$, and confirmed that it agrees with the general result Eq. (\ref{TR-tot}) in the limit $h \rightarrow -\infty$. Similar to the results of \cite{Cai:2018dkf} for the bispectrum, we have shown that the maximum values of $g_{NL}$ and $\tau_{NL}$ are obtained in the limit of an infinite sharp transition while in the opposite limit of a mild transition with $|h| \ll1$, much of $g_{NL}$ and $\tau_{NL}$ are washed out during the subsequent evolution towards the attractor phase. 
Finally, we have looked at the shape of trispectrum in various configurations.

There are a number of directions that the current studies can be extended. 
One interesting question is to consider the case where there are hierarchies between the modes such that some modes leave the horizon during the SR stage while long modes have already left the horizon during the USR stage. This case can be studied directly via in-in formalism while employing $\delta N$ formalism will be non-trivial now since the long and short modes  may not be treated as superhorizon modes simultaneously. 
Another natural question is to consider the higher order correlations $\langle \calR^n \rangle$ for $n>4$ in this setup and look at the corresponding dimensionless parameters similar to $g_{NL}$ and $\tau_{NL}$. However, 
for $n>4$,  more shapes and dimensionless parameters beyond   $g_{NL}$ and $\tau_{NL}$ have to be considered.

\vspace{0.7cm}
   
 {\bf Acknowledgments:}   We thank Xingang Chen, Mohammad Hossein Namjoo, Haidar Sheikhahmadi and Bahar Nikbakht 
 for insightful comments and discussions. 
 The work of H. F. is supported by INSF  of Iran under the grant  number 4046375. 
  
\vspace{.5cm}

\appendix
\section{Contributions from $H_3$}
\label{H3}

The analysis of in-in integrals associated to the right diagram of 
Fig. \ref{diagrams} with two vertices of ${\bf H_3}$  involves nested double time integrals. Here we outline the corresponding analysis yielding to the results quoted in section  \ref{H3TR}. 

Since two powers of ${\bf H_3}$ appear inside the in-in integral, we decompose  ${\bf H_3}$  as a combination of a quadratic power of $\calR$ and a single power of $\cal R$ as follows, 
\begin{equation}
  H_3 \supset  \alpha(\tau)  A(\tau) C(\tau) + \beta(\tau) 
  B(\tau) D(\tau) \, ,
\end{equation}
in which $\alpha(\tau)$ and $\beta(\tau)$ are  numerical functions of time in the cubic Hamiltonian. Furthermore,  
$A(\tau)$ and $B(\tau)$ are parts of $H_3$ which are quadratic in 
$\calR$ while $C(\tau)$ and $D(\tau)$ are linear in $\calR$. For example,
one may choose $A(\tau)=\mathcal{R}'(\tau)^2$ or $A(\tau)=\mathcal{R}(\tau)\mathcal{R}'(\tau)$ depending on the choice for terms that are contracted with external legs (the same holds for $B(\tau)$). Now if one chooses $A(\tau)=\mathcal{R}'(\tau)^2$ then 
$C(\tau)=\mathcal{R}(\tau)$ (the same holds for $D(\tau)$). Note that $A$ and $B$ may originate from both the gradient  and the time derivative 
terms while the contributions of the gradient terms proved to be suppressed. 

It is also convenient to  introduce the following definition, 
\begin{equation}
\begin{split}
&X_{k_1k_2}(\tau)\equiv \alpha(\tau)\mathcal{R}_{k_1}(\tau_0)\mathcal{R}_{k_2}(\tau_0)A(\tau)^*\\
&Y_{k_3k_4}(\tau)\equiv \beta(\tau)\mathcal{R}_{k_3}(\tau_0)\mathcal{R}_{k_4}(\tau_0)B^*(\tau) \, .
\end{split}
\end{equation}
 
We note that in the in-in integrals involving ${\bf H_3}$, it can appear 
at two times  at the left or right hand side of the operator, these contributions are  denoted by (2,0) or (0,2). Or,  it  can  appear once at each side simultaneously, which is  denoted by (1,1). Now using Eq. \eqref{dyson}  for the cubic Hamiltonian, one may write the combination of (2,0) and (0,2) as:
\subsection{(2,0) and (0,2) Contribution}
\begin{equation}
\begin{split}\label{02}
 &\left<\mathcal{R}_{k_1}(\tau_0)\mathcal{R}_{k_2}(\tau_0)\mathcal{R}_{k_3}(\tau_0)\mathcal{R}_{k_4}(\tau_0)\right>_{H_3,(0,2),(2,0)}=\\
&-\int_{-\infty}^{\tau_0}\int_{-\infty}^{\tau'}X_{k_1k_2}(\tau')Y_{k_3k_4}(\tau'')C(\tau')D^*(\tau'')d\tau''d\tau'\\
&-\int_{-\infty}^{\tau_0}\int_{-\infty}^{\tau'}X_{k_1k_2}(\tau'')Y_{k_3k_4}(\tau')D(\tau')C^*(\tau'')d\tau''d\tau'\\&-\int_{-\infty}^{\tau_0}\int_{-\infty}^{\tau'}X^*_{k_1k_2}(\tau')Y^*_{k_3k_4}(\tau'')C^*(\tau')D(\tau'')d\tau''d\tau'\\
&-\int_{-\infty}^{\tau_0}\int_{-\infty}^{\tau'}X^*_{k_1k_2}(\tau'')Y^*_{k_3k_4}(\tau')D^*(\tau')C(\tau'')d\tau''d\tau'
\end{split}
\end{equation}
Note that in the above expression, we have presented each term with its complex conjugate and the subscripts $(0,2),(2,0)$ denote that we have expanded the integral by two  time-ordered  or anti-time-ordered terms. 

On the other hand, for the case $(1,1)$ we have:
\subsection{(1,1) Contribution}
\begin{equation}
\begin{split}\label{11}
&\left<\mathcal{R}_{k_1}(\tau_0)\mathcal{R}_{k_2}(\tau_0)\mathcal{R}_{k_3}(\tau_0)\mathcal{R}_{k_4}(\tau_0)\right>_{H_3,(1,1)} = \\
&\int_{-\infty}^{\tau_0}\int_{-\infty}^{\tau_0}A(\tau')C(\tau')\mathcal{R}^*_{k_1}(\tau_0)\mathcal{R}^*_{k_2}(\tau_0)\mathcal{R}_{k_3}(\tau_0)\mathcal{R}_{k_4}(\tau_0)B^*(\tau'')D^*(\tau'')d\tau'd\tau''+ \mathrm{C.C}=\\&\int_{-\infty}^{\tau_0}\int_{-\infty}^{\tau_0}X_{k_1k_2}^*(\tau')Y_{k_3k_4}(\tau'')C(\tau')D^*(\tau'')d\tau'd\tau''\\&+\int_{-\infty}^{\tau_0}\int_{-\infty}^{\tau_0}X_{k_1k_2}(\tau')Y^*_{k_3k_4}(\tau'')C^*(\tau')D(\tau'')d\tau'd\tau''.
    \end{split}
\end{equation}
 Now we note that, 
\begin{equation}
\begin{split}
\label{25}
    &\int_{-\infty}^{\tau_0}\int_{-\infty}^{\tau_0} X_{k_1k_2}^*(\tau')Y_{k_3k_4}(\tau'')C(\tau')D^*(\tau'')d\tau''d\tau'=\\
    &\int_{-\infty}^\tau\int_{-\infty}^{\tau'} X_{k_1k_2}^*(\tau')Y_{k_3,k_4}(\tau'')C(\tau')D^*(\tau'')d\tau''d\tau'\\
    &+\int_{-\infty}^\tau\int_{\tau'}^{\tau_0}X_{k_1,k_2}^*(\tau')Y_{k_3,k_4}(\tau'')C(\tau')D^*(\tau'')d\tau''d\tau'
    \end{split}
\end{equation}
By changing the order of integrals in the second integration in the above equation we may write,
\begin{equation}
\begin{split}
\label{change}
    &\int_{-\infty}^{\tau_0}\int_{\tau'}^{\tau_0}X_{k_1k_2}^*(\tau')Y_{k_3k_4}(\tau'')C(\tau')D^*(\tau'')d\tau''d\tau'\\&=\int_{-\infty}^{\tau_0}\int_{-\infty}^{\tau''}X_{k_1,k_2}^*(\tau')Y_{k_3k_4}(\tau'')C(\tau')D^*(\tau'')d\tau'd\tau''\\&=\int_{-\infty}^{\tau_0}\int_{-\infty}^{\tau'}X_{k_1,k_2}^*(\tau'')Y_{k_3,k_4}(\tau')C(\tau'')D^*(\tau')d\tau''d\tau' \, ,
    \end{split}
\end{equation}
where in the last equality we have renamed $\tau''$ by $\tau'$ and vice versa. 

As the integrals coming from $(1,1)$ have a different sign compared to $(0,2)$ and $(2,0)$, we may combine the above integral with the last one   in \eqref{02} and write,
\begin{equation}
\label{1}
\text{Last in \eqref{02}+\eqref{change}}=2 i\int_{-\infty}^{\tau_0}\int_{-\infty}^{\tau'} \mathrm{Im}(Y_{k_3 k_4}(\tau'))D^*(\tau')X^*_{k_1k_2}(\tau'')C(\tau'')d\tau''d\tau' \, .
\end{equation}
In the same manner, the first integral in \eqref{25} can be combined with the first one in \eqref{02}, yielding, 
\begin{equation}
\label{2}
  \text{First in \eqref{02}+First in \eqref{25}}=-2 i\int_{-\infty}^{\tau_0}\int_{-\infty}^{\tau'} \mathrm{Im}(X_{k_1 k_2}(\tau'))D^*(\tau'')Y_{k_3 k_4}(\tau'')C(\tau')d\tau''d\tau' \, .
\end{equation}
Finally we combine the second integral in \eqref{11} with the second and third integrals in \eqref{02} and write:
\begin{equation}
\label{3}
  \text{Second in \eqref{02}+Second in \eqref{11}}=-2 i\int_{-\infty}^{\tau_0}\int_{-\infty}^{\tau'} \mathrm{Im}(Y_{k_3 k_4}(\tau'))D(\tau')X_{k_1 k_2}(\tau'')C^*(\tau'')d\tau''d\tau' \, ,
\end{equation}
\begin{equation}
\label{4}
  \text{third in \eqref{02}+Second in \eqref{11}}=2i\int_{-\infty}^{\tau_0}\int_{-\infty}^{\tau'} \mathrm{Im} (X_{k_1 k_2}(\tau'))D(\tau'')Y_{k_3 k_4}(\tau'')C^*(\tau')d\tau''d\tau' \, .
\end{equation}
Combining \eqref{1} with \eqref{3}  yields, 
\begin{equation}
\label{firH3}
    \text{\eqref{1}+\eqref{3}}=4\times{\text{perm}}\int_{-\infty}^{\tau_0}\int_{-\infty}^{\tau'} \mathrm{Im} (Y_{k_3 k_4}(\tau')) \mathrm{Im} \big(X_{k_1 k_2}(\tau'')C^*(\tau'')D(\tau') \big),
\end{equation}
where $\text{perm}$ denotes the possible numbers of ways that one can construct $A(\tau)$ or $B(\tau)$. This factor differs case by case and depending on what one chooses for $A(\tau)$ and $B(\tau)$. 

Similarly, one can combine \eqref{2} with \eqref{4} to obtain,
\begin{equation}
\label{secH3}
    \text{\eqref{2}+\eqref{4}}=4\times{\text{perm}}\int_{-\infty}^{\tau_0}\int_{-\infty}^{\tau'} \mathrm{Im} (Y_{k_3 k_4}(\tau')) \mathrm{Im} \big(X_{k_1 k_2}(\tau'')C^*(\tau'')D(\tau') \big) \, .
\end{equation}

Combining \eqref{firH3} with \eqref{secH3} and considering all permutations, we finally obtain, 
\begin{equation}
\label{H3tot}
\begin{split}
  & \left<\mathcal{R}_{k_1}(\tau_0)\mathcal{R}_{k_2}(\tau_0)\mathcal{R}_{k_3}(\tau_0)\mathcal{R}_{k_4}(\tau_0)\right>_{H_3} =4\times{\text{perm}}\int_{-\infty}^{\tau_0}\int_{-\infty}^{\tau'} \mathrm{Im} (Y_{k_3 k_4}(\tau')) \mathrm{Im} (X_{k_1 k_2}(\tau'')C^*(\tau'')D(\tau')) \\ 
  &\hspace{6cm} +23 \text{permutations} \, .
   \end{split}
\end{equation}

Now, we make a list in which different options for $A$ and $B$ and their complements in Hamiltonian are presented. Specifically,  
\begin{itemize}
    \item I (Time derivative-Time derivative)
    
    \begin{itemize}
    \item I  :  $A=\mathcal{R}\mathcal{R}'$ , $B=\mathcal{R}\mathcal{R}'$, $\text{perm}=4$
     \item II : $A=\mathcal{R}\mathcal{R}'$ , $B=\mathcal{R}'^2$, $\text{perm}=2$
      \item IV: $A=\mathcal{R}'^2$ , $B=\mathcal{R}\mathcal{R}'$, $\text{perm}=2$
     \item III: $A=\mathcal{R}'^2$ , $B=\mathcal{R}'^2$, $\text{perm}=1$
    \end{itemize}
     \item II (Time derivative-Spatial Gradient)
     \begin{itemize}
    \item I: $A=\mathcal{R}\mathcal{R}'$ , $B=(\partial\mathcal{R})^2$, $\text{perm}=2$
     \item II: $A=\mathcal{R}\mathcal{R}'$ , $B=\calR\partial_i R$, $\text{perm}=4$ 
     \item III:  $A=\mathcal{R}'^2$ , $B=\calR\partial_i \calR$, $\text{perm}=2$
     \item IV: $A=\mathcal{R}'^2$ , $B=(\partial\mathcal{R})^2$, $\text{perm}=1$
     \end{itemize}
      \item III (Saptial Gradient-Spatial Gradient)
     \begin{itemize}
    \item I: $A=\calR\partial_i\mathcal{R}$,  $B=\calR\partial_i\mathcal{R}$ , $\text{perm}=4$ 
     \item II:  $A=\calR\partial_i\mathcal{R}$, ,$B=(\partial\mathcal{R})^2$, $\text{perm}=2$
      \item III:  $A=(\partial\mathcal{R})^2$,  $B=(\partial\mathcal{R})^2$, $\text{perm}=1$
      \item IV: $A=(\partial\mathcal{R})^2$,  $B=\calR\partial_i\mathcal{R}$,  $\text{perm}=2$
    \end{itemize}
\end{itemize}

\section{Contributions from non-linear terms }
\label{NLterms}

Here we outline the contributions from the non-linear relations between 
$\pi$ and $\calR$ which have been used in section \ref{sharp-tran} for the trispectrum in the setup with an infinitely sharp transition.

The non-linear relation between $\pi$ and $\calR$ to leading orders in slow-roll parameter  is given in Eq. (\ref{nonlinear2}). We consider the contributions of each non-linear terms separately. For this purpose note that   the following relations for the derivatives of $\pi$ at the end of USR can be used \cite{Akhshik:2015nfa}, 
\begin{equation}
    \dot\pi=-3 \calR \, , \quad \quad   \ddot\pi=-9H \calR \, .
\end{equation}

Moreover as we need the Fourier transformation of the non-linear terms, we make use of the following convolution integrals,
\begin{equation}
   ( \pi \dot\pi)_{\bfk_1}=\int  \frac{d^3q}{(2\pi)^3} \pi_\bfq\dot\pi_{\bfk-\bfq} ,
\end{equation}
\begin{equation}
   ( \pi^2 \ddot\pi)_{\bfk_1}=\int\int  \frac{d^3qd^3q_1}{(2\pi)^6} \pi_{\bfq_1}\pi_{\bfq-\bfq_1}\ddot\pi_{\bfk_1-\bfq} \, ,
\end{equation}
\begin{equation}
   ( \dot\pi^2 \pi)_{\bfk_1}=\int\int  \frac{d^3qd^3q_1}{(2\pi)^6}
   \dot\pi_{\bfq_1}\dot\pi_{\bfq-\bfq_1}\pi_{\bfk_1-\bfq} \, .
\end{equation}

Let us start with the quadratic term $H \pi \dot \pi$. Since we need an even number of fields to obtain a non-zero correlator, we require one power of
${\bf H_3}$. There are two different contributions,      
\begin{itemize}
    \item{$H \pi\dot\pi$+${\bf H_3}$}
    \begin{equation}
\begin{split}
  &\left<\mathcal{R}_{k_1}(\tau_0)\mathcal{R}_{k_2}(\tau_0)\mathcal{R}_{k_3}(\tau_0)\mathcal{R}_{k_4}(\tau_0)\right>_{ \pi \dot \pi_{(1)}}\\& =-6\eta \mathrm{Im}\Big<\int\frac{\mathcal{R}_{\bfk_1-{\bf l}}\mathcal{R}_{\bf l} d^3 {\bf l}}{(2\pi)^3 } \mathcal{R}_{k_2}\mathcal{R}_{k_3}\mathcal{R}_{k_4}\\&\times\int\epsilon a^2 \mathcal{R}_{q_1} \calR'_{q_2} \calR'_{q_3}(2\pi)^3\delta^3(q_1+q_2+q_3)d\tau\frac{d^3q_1d^3q_2d^3q_3}{((2\pi)^3)^3})\Big>\\&=-24 \mathrm{Im} \Big[\mathcal{R}_{k_{12}}(\tau_0)|\mathcal{R}_{k_2}(\tau_0)|^2\mathcal{R}_{k_3}^*(\tau_0)\mathcal{R}_{k_4}(\tau_0)\int_{-\infty}^{\tau_0} \epsilon a^2 \eta \calR^*_{k_{12}}(\tau')\calR'^*_{k_{3}}(\tau')R'^*_{k_{4}}(\tau') \Big]d\tau'\\&+11 \text{perms.} \, .
  \end{split}
\end{equation}
The  other possible contribution is the case in which the non-linear term is contracted with the time derive contributions in the Hamiltonian. This case is easily obtained by replacing ${k_3\leftrightarrow k_{12}}$ in the above, yielding
\begin{equation}
\begin{split}
  &\left<\mathcal{R}_{k_1}(\tau_0)\mathcal{R}_{k_2}(\tau_0)\mathcal{R}_{k_3}(\tau_0)\mathcal{R}_{k_4}(\tau_0)\right>_{  \pi \dot \pi_{(2)}}\\&=-24 \mathrm{Im} \Big[\mathcal{R}_{k_{3}}(\tau_0) |\mathcal{R}_{k_2}(\tau_0)|^2 \mathcal{R}_{k_{12}}^*(\tau_0)\mathcal{R}_{k_4}(\tau_0)\int_{-\infty}^{\tau_0} \epsilon a^2 \eta  \calR^*_{k_{3}}(\tau') \calR'^*_{k_{12}}(\tau') \calR'^*_{k_{4}}(\tau') 
  \Big] d\tau'\\&+23 \, \text{perms.}
  \end{split}
\end{equation}

 \item{$H \pi\dot\pi$+$H \pi\dot\pi$}\\\\
 Now consider the case when two non-linear terms from 
 $H \pi \dot \pi$ contribute jointly. The analysis is simpler now as we have even numbers of field in the correlator and there is no need for ${\bf H}_3$.
The result is given as,
 \begin{equation}
 \begin{split}
     &\left<\mathcal{R}_{k_1}(\tau_0)\mathcal{R}_{k_2}(\tau_0)\mathcal{R}_{k_3}(\tau_0)\mathcal{R}_{k_4}(\tau_0)\right>_{ \pi \dot \pi \pi \dot \pi}=9 \times 2|\mathcal{R}_{k_{24}}(\tau_0)|^2|\mathcal{R}_{k_{3}}(\tau_0)|^2|\mathcal{R}_{k_{4}}(\tau_0)|^2\\&+23 \text{Permutations}
     \end{split}
 \end{equation}

\item{$\frac{H}{2} \ddot\pi\pi^2$ or $H \dot\pi^2\pi$}\\\\
 Finally, we have the contributions from non-linear terms $\frac{H}{2} \ddot\pi\pi^2$ or $H \dot\pi^2\pi$.  Similar to the above case, there are even number of fields inside the correlator and there is no need for ${\bf H}_3$. We obtain, 
  \begin{equation}
  \begin{split}
     &\left<\mathcal{R}_{k_1}(\tau_0)\mathcal{R}_{k_2}(\tau_0)\mathcal{R}_{k_3}(\tau_0)\mathcal{R}_{k_4}(\tau_0)\right>_{ \pi \dot \pi^2}=9 \times 6|\mathcal{R}_{k_{24}}(\tau_0)|^2|\mathcal{R}_{k_{3}}(\tau_0)|^2|\mathcal{R}_{k_{4}}(\tau_0)|^2\\&+3 \text{Permutations} \, ,
     \end{split}
 \end{equation}
 and, 
 \begin{equation}
 \begin{split}
    & \left<\mathcal{R}_{k_1}(\tau_0)\mathcal{R}_{k_2}(\tau_0)\mathcal{R}_{k_3}(\tau_0)\mathcal{R}_{k_4}(\tau_0)\right>_{ \pi^2 \ddot \pi}=\frac{9}{2} \times 6|\mathcal{R}_{k_{2}}(\tau_0)|^2|\mathcal{R}_{k_{3}}(\tau_0)|^2|\mathcal{R}_{k_{4}}(\tau_0)|^2+\\&3 \text{Permutations} \, .
     \end{split}
 \end{equation}
\end{itemize}
The above results are presented in Eqs. (\ref{PiR2a}),  (\ref{PiR2b}), (\ref{PiR3a}) and (\ref{PiR3b}).

 \bigskip


\bibliography{trispectrum}{}

\bibliographystyle{JHEP}

\end{document}